\documentclass[a4paper]{article}

\usepackage{INTERSPEECH2020}

\usepackage{booktabs}
\usepackage{dcolumn}
\usepackage{soul}
\usepackage{multirow}
\usepackage{color}
\usepackage{romannum}
\DeclareMathSymbol{:}{\mathord}{operators}{"3A}
\newcolumntype{L}[1]{>{\raggedright\let\newline\\\arraybackslash\hspace{0pt}}m{#1}}
\newcolumntype{C}[1]{>{\centering\let\newline\\\arraybackslash\hspace{0pt}}m{#1}}
\newcolumntype{R}[1]{>{\raggedleft\let\newline\\\arraybackslash\hspace{0pt}}m{#1}}
\title{Multimodal Speech Emotion Recognition using Cross Attnention\\with Aligned Audio and Text}
\name{Yoonhyung Lee, Seunghyun Yoon, Kyomin Jung}
\address{
  Department of Electrical and Computer Engineering,\\
  Seoul National University, Seoul, South Korea
  }
\email{cpi1234@snu.ac.kr, mysmilesh@snu.ac.kr, kjung@snu.ac.kr}

\begin{document}

\maketitle

\begin{abstract}
In this paper, we propose a novel speech emotion recognition model called Cross Attention Network (CAN) that uses aligned audio and text signals as inputs.
It is inspired by the fact that humans recognize speech as a combination of simultaneously produced acoustic and textual signals.
First, our method segments the audio and the underlying text signals into equal number of steps in an aligned way so that the same time steps of the sequential signals cover the same time span in the signals.
Together with this technique, we apply the cross attention to aggregate the sequential information from the aligned signals.
In the cross attention, each modality is aggregated independently by applying the global attention mechanism onto each modality.
Then, the attention weights of each modality are applied directly to the other modality in a crossed way, so that the CAN gathers the audio and text information from the same time steps
based on each modality.
In the experiments conducted on the standard IEMOCAP dataset, our model outperforms the state-of-the-art systems by 2.66\% and 3.18\% relatively in terms of the weighted and unweighted accuracy.
\end{abstract}
\noindent\textbf{Index Terms}: speech emotion recognition, multimodal learning, deep learning, attention mechanism

\section{Introduction}
% Background and importance of the Speech Emotion Recognition
In developing human-computer interaction systems, Speech Emotion Recognition (SER) technology is considered as an essential element to provide proper response depending on a user's emotional state \cite{kolakowska2014emotion}.
Many machine learning models have been built for SER, in which the models are trained to predict an emotion among the candidates such as happy, sad, angry, or neutral for a given speech~\cite{nwe2003speech, chavhan2010speech, mao2014learning, mirsamadi2017automatic}.
Recently, researchers have adopted multimodal approaches in SER considering that emotions can be expressed in various ways such as facial expressions, gestures, texts, or speech~\cite{castellano2008emotion, yoon2020attentive}.
In particular, the text modality has been frequently used in addition to the speech in many SER studies, because human speech inherently consists of the acoustic features and the linguistic contents that can be expressed using text~\cite{yoon2019speech, xu2019learning}.

% Previous studies
The major issue in SER using both the audio and text modalities is how to extract and combine the information that each audio and text carries.
For example, if someone says, \emph{``Thank you for being with me."} in a very calm voice, the emotional information is contained mostly in the linguistic contents while it sounds neutral based on the acoustic features.
Previous studies have approached this issue by designing their models to encode the audio and text independently and fuse the results using attention mechanisms, which help their models effectively capture the locally salient regions from given signals.
In these attention mechanisms, the separately encoded audio and text information operated as each other's query and key-value pair.
Yoon et al.~\cite{yoon2019speech} used the last hidden state of a recurrent modality encoder as a query and used the other encoded modality as a key-value pair in the attention mechanism.
In another research, Xu et al.~\cite{xu2019learning} designed their model to learn the alignment between the audio and text by itself from the attention mechanism.

% Locate a gap in the existing literature
However, letting the model learn the complex interaction between the different modalities without any constraints can make the training more difficult.
Using the last hidden state of a recurrent encoder as a query as in \cite{yoon2019speech} can lead to temporal information loss in the attention as pointed out in \cite{mirsamadi2017automatic}.
% \textcolor{red}{Learning the alignment between the audio and text by a model itself as in \cite{xu2019learning} is a challenging task.
% Actually, learning the alignment between the audio and the text is one of the difficult problems in the speech recognition or the speech synthesis \cite{raffel2017online, battenberg2019location}, where much larger dataset is required in the tasks compared to the SER.}
% \textcolor{red}{Also, for a model, learning the alignment between the audio and text from the attention mechanism as in \cite{xu2019learning} is a challenging task as pointed out in \cite{raffel2017online, battenberg2019location}.}
Besides, learning the alignment between the audio and text signals relying on the attention mechanism as in \cite{xu2019learning} is a challenging task unless additional prior knowledge is provided as in \cite{raffel2017online, battenberg2019location}.
% \textcolor{red}{ 
% \input{ForcedAlignment.tex}
% The models learned the attention between audio and text modality.
% Yoon et al.~\cite{yoon2019speech} use a summarized vector obtained from a single modality as a query in the attention mechanism.
% However, considering that it is important to effectively capture the salient parts from the input signals in SER \cite{mao2014learning, mirsamadi2017automatic, bertero2017first}, using the summarized vector in attention can lead to information loss.
% In another research, Xu et al.~\cite{xu2019learning} design their model to learn the alignment between the audio and text automatically by employing the attention mechanism.
% However, the audio and text data consisting of different time steps with different time resolutions make it difficult for the model to learn the alignment without any constraints.
% Actually, learning the alignment between audio and text is a challenging task in speech recognition or speech synthesis, where even much more data is used compared to the SER~\cite{raffel2017online, battenberg2019location}.
% Above all, independently encoding the audio and text first, and then fusing them using attention mechanism does not reflect the human recognition of the spoken language.
% When humans recognize the spoken language, textual and acoustic information are recognized simultaneously.
% and the salient parts for SER are different depending on the modalities.
% }

% Explain this paper
\begin{table}
  \caption{An example of the alignment information provided in the IEMOCAP dataset. The numbers in the table represent the timing when the each uttering of the words begins and ends in the speech. The values are expressed in 10 milliseconds unit. $\langle s \rangle$, $\langle /s \rangle$, and $\langle sil \rangle$ are special tokens meaning the start and end of a sentence, and the silence.}
  \label{tab:alignment}
  \centering
  \begin{tabular}
    {C{0.25\columnwidth}C{0.25\columnwidth}C{0.3\columnwidth}}
    \toprule
    start & end & word \\
    \midrule
    0   &   51  &   $\langle s \rangle$ \\
    52  &   75  &   i \\    
    76  &   88  &   like \\
    89  &   140 &   $\langle sil \rangle$  \\
    141 &   143 &   apple   \\
    144 &   177 &   $\langle /s \rangle$   \\
    \bottomrule
    \end{tabular}
\end{table}
To overcome these limitations, we propose a novel SER model called Cross Attention Network (CAN) that effectively combines the information obtained from aligned audio and text signals.
Inspired by how humans recognize speech, we design our model to regard the audio and text as temporarily aligned signals.
In the CAN, each audio and text input is separately encoded through its own recurrent encoder.
Then, the hidden states obtained from each encoder are independently aggregated by applying the global attention mechanism onto each modality.
Furthermore, the attention weights extracted from each modality are directly applied to each other's hidden states in a crossed way, so that the information at the same time steps is aggregated with the same weights.
% The crossed aggregation is only possible when the audio and text signals have the same time steps.

% In order to make the cross attention works properly, we propose an aligned segmentation that segments the audio and text signals to have the same time steps \textcolor{red}{with the same time resolution.}
% \textcolor{red}{In order to make the cross attention works properly, we propose an aligned segmentation, where the audio and text signals are segmented into the same number of parts.
% In the aligned segmentation, the start and the end of each word are also considered, which are provided in the alignment information as described in the example of Table~\ref{tab:alignment}.
% In other words, the audio and text data are segmented to have the same time steps with the same time resolution.}
In order to make the cross attention work properly, we propose an aligned segmentation technique that divides each audio and text signal into the same number of parts in an aligned way.
In the aligned segmentation technique, the text signal is segmented into words.
Following the text, the audio signal is segmented using alignment information as shown in Table~\ref{tab:alignment}, where the start- and end-time for each word are used to determine the partitioning points in the audio signal.
The aligned segmentation technique enables our model to successfully combine the information from the aligned audio and text signals without having to learn the complex attention between different modalities as in the previous works.

% What we saw in this study
To evaluate the performance of the proposed method, we conduct experiments on the IEMOCAP dataset.
Firstly, we compare the CAN with other state-of-the-art SER models that use additional text modality.
The results show that our model outperforms the other models in both weighted and unweighted accuracy by 2.66\% and 3.18\% relatively.
% Experimental results indicate that the CAN outperforms the state-of-the-art model for the SER task.
Furthermore, ablation studies are conducted to see the actual effectiveness of the components such as aligned segmentation, stop-gradient operator, and additional loss.
In the ablation studies, we observe the independent contribution of each component for improving the model performance.
% Experimental results indicate that the CAN outperforms the other models including multi-modal models.
% In other experiments, we show that the aligned segmentation scheme is also helpful when it is used in other multi-modal models and the aligned segmentation policy is really effective compared to equal or random segmentation policies.
% To the best of our knowledge, this is the first study that aggregating the information from the aligned audio and text actually bring the performance gain in the SER.

\section{Related work}
After the classical machine learning models such as the hidden markov model or the support vector machine \cite{nwe2003speech, chavhan2010speech}, models using neural networks have been actively studied in Speech Emotion Recognition (SER).
To improve the model performance, researchers proposed various methods to effectively capture the locally salient regions over the time axis from a given speech.
% For the models, it was important to design the models to effectively capture the locally salient regions over the time axis from the speech.
Bertero et al. \cite{bertero2017first} proposed a model consisting of the convolutional neural network (CNN) that captures local information from given acoustic feature frames.
Mirsamadi et al. \cite{mirsamadi2017automatic} used the global attention mechanism to make their model learn where to attend to capture the locally salient features.
Sahoo et al. \cite{sahoo2019segment} proposed to train a CNN-based model with audio segments that are segmented from an utterance with equal length, which improved the model by forcing it to learn to capture the locally salient emotional features in a more elaborated manner.

Recently, multimodal models that use the audio and text together for SER have attracted much attention \cite{yoon2019speech, xu2019learning, sebastian2019fusion, liang2019cross}.
Since the audio and text signals contain different information, it has been a major issue of how to design the models to effectively extract information from each modality and combine them.
In the previous studies, attention mechanisms were frequently used to combine the information \cite{yoon2019speech, xu2019learning}, where the hidden states obtained separately from the audio and text signals were used as each other's query or key-value pair.
The attention mechanisms were expected to help their models learn to combine the information of each modality by themselves.
However, none of these studies used proper constraints of prior knowledge to ease the difficulty of learning the complex interaction between the audio and text signals.

\section{Methodology}
\label{section:algorithm}
In this section, we propose a novel Speech Emotion Recognition (SER) model called Cross Attention Network (CAN).
First, we explain a preprocessing of the text and audio data, which is necessary for the CAN to work properly.
The purpose of the preprocessing is to make the text and audio have the same number of time steps while the same time steps of the sequential signals cover the same time span.
Then the CAN is explained, which is a model utilizing the cross attention mechanism that enables the CAN to focus on the salient features of the aligned text and audio signals with a different perspective of each modality.

\subsection{Data preprocessing}
\subsubsection{Text data}
In this study, we consider a text input as a word sequence, so the text input is represented as $X=\{x_1, x_2, ..., x_L\},~X\in \mathbb{R}^{L\times V}$, where $L$ is the number of words, $V$ is the size of the vocabulary, and each $x_i$ is a one-hot vector representing the corresponding word.
Then, $\textbf{E}^{(T)}\in \mathbb{R}^{L\times D_e}$, an embedded text input, is obtained after the $X$ passes through a trainable Glove embedding layer \cite{pennington2014glove}, where $D_e$ is the dimension of the embedding layer.

\subsubsection{Audio data}
\label{subsection:audioprocessing}
Let $Y=\{y_1, y_2, ..., y_T\},\;Y\in \mathbb{R}^{T}$ be the 1-dimensional audio data and $D=\{d_1, d_2, ..., d_L\}$ be its alignment information, where $T$ is the audio length and each $d_i=(s_i, e_i)$ represents the start and the end of each word.
To prevent information loss about the correlation, we make the neighboring $d_i$s have 10\% overlap.

Using the $Y$ and $D$, we obtain a segmented audio data $\textbf{E}^{(A)}\in \mathbb{R}^{L\times (T'\times D_f)}$; the audio $Y$ is first segmented into audio segments $Y'=\{y_{s_1:e_1},y_{s_2:e_2},...,\;y_{s_L:e_L}\}$, and then each segment is converted into a MFCC feature and stacked into the $\textbf{E}^{(A)}$ with zero-padding.
Here, $D_f$ is the number of the MFCC coefficients and $T'$ is the length of the longest MFCC.

% Using the $Y$ and $D$, we obtain a segmented audio data $\textbf{E}^{(A)}\in \mathbb{R}^{L\times (T'\times D_f)}$ as follows:
% \begin{enumerate}
%   \item Segment the audio $Y$ into the audio segments $Y'=\{y_{s_1:e_1},y_{s_2:e_2},...,\;y_{s_L:e_L}\}$.
%   \item Obtain Mel-Frequency Cepstral Coefficients (MFCC) features $\textbf{E}^{(A)}_i\in \mathbb{R}^{(e_i-s_i) \times D_f}$ from each audio segment.
%   \item Stack the MFCC features into one audio signal with zero-padding.
% \end{enumerate}
% Here, $D_f$ is the number of the MFCC coefficients and $T'$ is the length of the longest MFCC.
% In this way, each audio segment $\textbf{E}^{(A)}_i$ is aligned with the corresponding word; thus, the overall text data $\textbf{E}^{(T)}$ and the audio data $\textbf{E}^{(A)}$ have the same time resolution.

%   \item Obtain Mel-Frequency Cepstral Coefficients (MFCC) features $\textbf{E}^{(A)}_i\in \mathbb{R}^{(e_i-s_i) \times D_f}$ from each audio segment.
%   \item Stack the MFCC features into one audio signal with zero-padding.
% \end{enumerate}
% Here, $D_f$ is the number of the MFCC coefficients and $T'$ is the length of the longest MFCC.
% In this way, each audio segment $\textbf{E}^{(A)}_i$ is aligned with the corresponding word; thus, the overall text data $\textbf{E}^{(T)}$ and the audio data $\textbf{E}^{(A)}$ have the same time resolution.

\subsection{Model architecture}
\begin{figure}[t]
    \centering
    \includegraphics[scale=0.20]{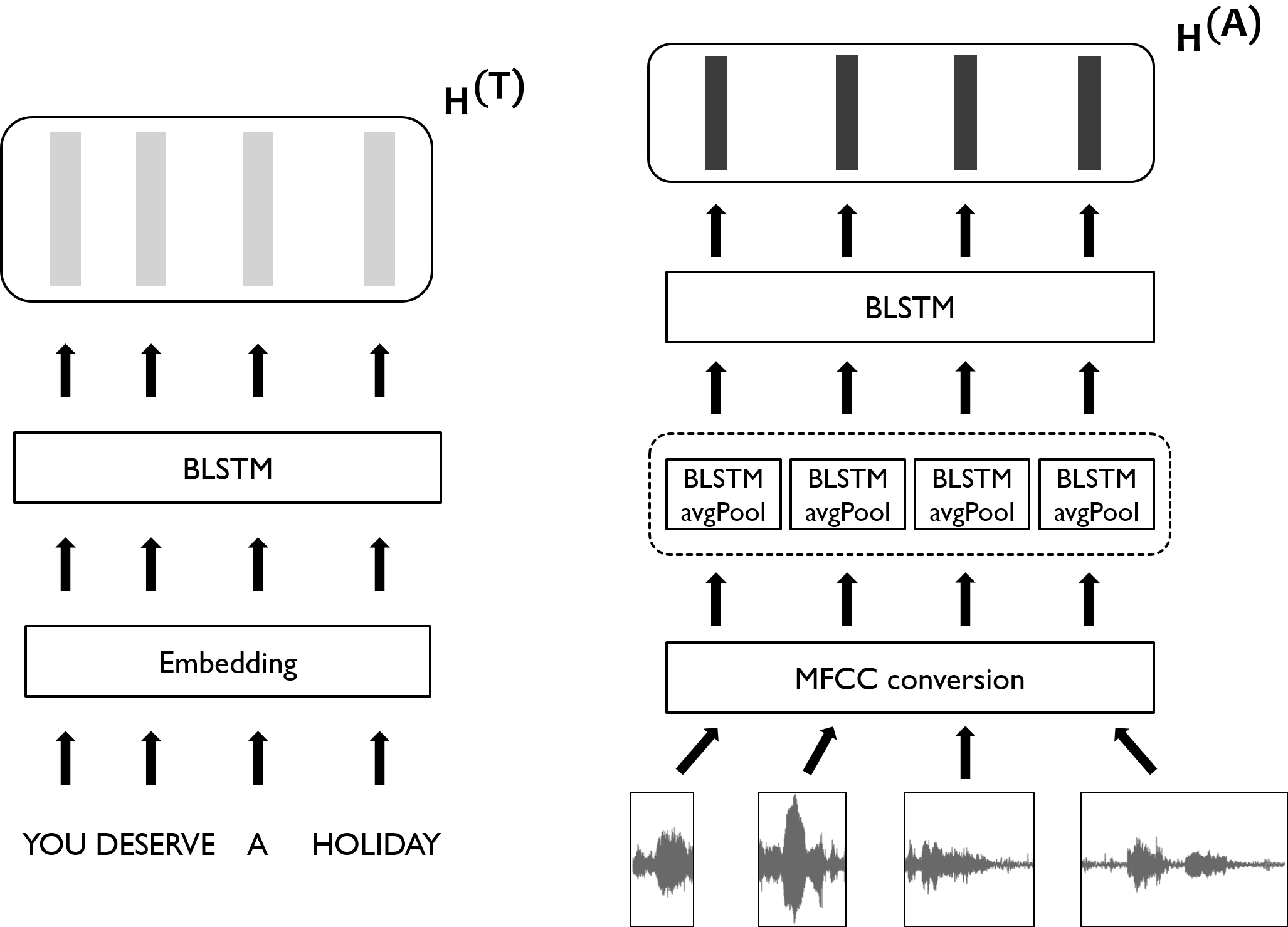}
    \caption{Text encoder and Audio encoder. The BLSTM modules inside the dotted box share their weights. For better understanding, we represent the audio input using raw waveforms but we convert the wavs into MFCC segments in advance and use them as audio inputs.}
    \label{fig:encoders}
\end{figure}
\subsubsection{Text encoder}
The embedded text data $\textbf{E}^{(T)}$ is fed into the text encoder consisting of the bidirectional long short-term memory (BLSTM) \cite{hochreiter1997long} as represented at the left side of Figure \ref{fig:encoders}, which leads to the hidden states $\textbf{H}^{(T)}\in\;\mathbb{R}^{L\times D_h}$ obtained from the equations below:
\begin{align}
 & \overrightarrow{h_i}=f_{\theta}(\overrightarrow{h_{i-1}},~ \textbf{E}^{(T)}_i), \\[10pt]
 & \overleftarrow{h_i}=f'_{\theta}(\overleftarrow{h_{i+1}},~ \textbf{E}^{(T)}_i), \\[10pt]
 & \textbf{H}^{(T)}=\{[\overrightarrow{h_1};\overleftarrow{h_1}],~ [\overrightarrow{h_2};\overleftarrow{h_2}],~...,~ [\overrightarrow{h_L};\overleftarrow{h_L}]\},
\end{align}
where $f_\theta$, $f'_\theta$ are the forward and backward LSTMs having $D_h$ hidden units with parameter $\theta$.
Additionally, $h_i$ represents the hidden state at i-th time step and $\textbf{E}^{(T)}_i$ represents the i-th embedded word vector of the text data.

\subsubsection{Audio encoder}
The audio encoder consists of two bidirectional LSTM layers as represented at the right side of Figure \ref{fig:encoders}.
The bottom LSTM layer encodes each MFCC segment $\textbf{E}^{(A)}_i\in \mathbb{R}^{T'\times D_f}$ independently and outputs a vector from each segment using average pooling.
The BLSTM modules inside the dotted box in Figure \ref{fig:encoders} share their weights.
The upper LSTM layer encodes the audio features obtained from the bottom layer and outputs the hidden states $\textbf{H}^{(A)}\in \mathbb{R}^{L\times D_h}$, which has the same time steps $L$ with the $\textbf{H}^{(T)}$.

% \subsection{Aligned segmentation}
% In order to use attention weights obtained from one modality to aggregate the hidden states of the other modality, the time steps of two hidden states should be same.
% Therefore, we first segment the audio and two step 
% Imitating the way human recognize spoken language, we use aligned audio and text data based on words to segment the audio. The format of the alignment information is shown in Figure 1.
% First, we remove special `silence' tokens and distribute its time span to the previous and the next words.
% Second, in order to make the segments be overlapped, for every word, we separate the last 25\% of its previous word's time span and the first 25\% of it's next word's time span and concatenate them with the middle word.
% As a result, the audio and text have same time steps and we can consider that the time steps represent real time because they are timely aligned.

\subsubsection{Cross attention}
\begin{figure}[t]
    \centering
    \includegraphics[scale=0.20]{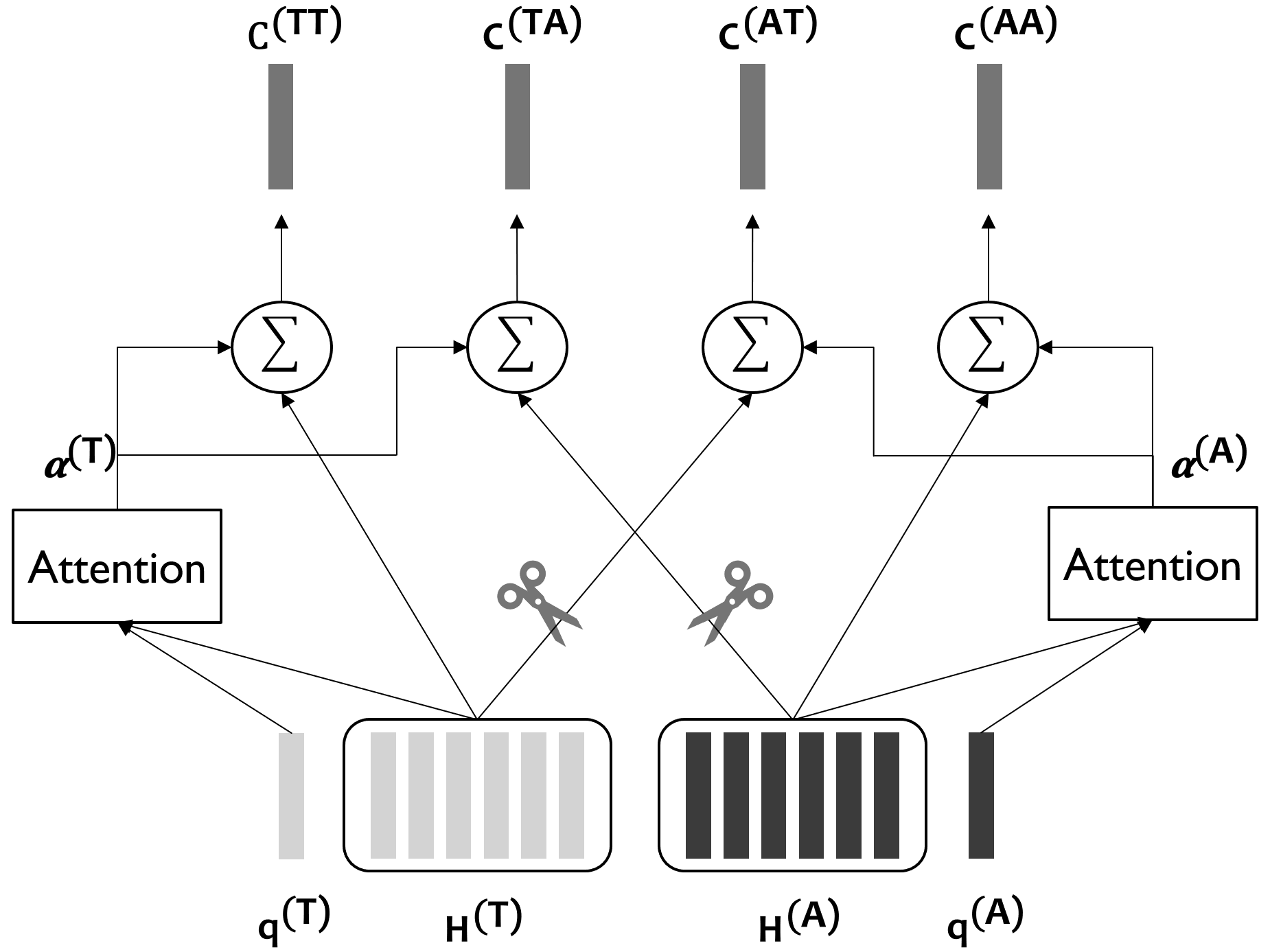}
    \caption{Cross Attention Network. The scissors represent the stop-gradient operator that cuts the gradient flow during back propagation.}
    \label{fig:can}
\end{figure}
In the cross attention, attention weights obtained from one modality are used to aggregate the other modality as shown in Figure \ref{fig:can}, while conforming to the constraint that the audio and text are temporarily aligned.
Since the salient regions can be different depending on what modality the prediction is based on, the aggregation of the modalities happens twice based on each modality in the cross attention as follows:
\begin{gather}
	\alpha^{(T)}_i=\dfrac{\text{exp}({~(\textbf{q}^{(T)})}^\intercal~\textbf{H}^{(T)}_{i}~)}{\sum_{j} \text{exp}({~(\textbf{q}^{(T)})}^\intercal~\textbf{H}^{(T)}_{j}~)},~~~(i=1,...,L),
    \\[10pt]
    \alpha^{(A)}_i=\dfrac{\text{exp}({~(\textbf{q}^{(A)})}^\intercal~\textbf{H}^{(A)}_{i}~)}{\sum_{j} \text{exp}({~(\textbf{q}^{(A)})}^\intercal~\textbf{H}^{(A)}_{j}~)},~~~(i=1,...,L),
\end{gather}
\begin{align}
    \textbf{c}^{(TT)}&={\sum_{i}} \alpha^{(T)}_i~{\textbf{H}^{(T)}_i},
    \\
    \textbf{c}^{(TA)}&={\sum_{i}} \mathbf{sg}(\alpha^{(T)}_i)~{\textbf{H}^{(A)}_i},
    \\
    \textbf{c}^{(AA)}&={\sum_{i}} \alpha^{(A)}_i~{\textbf{H}^{(A)}_i},
    \\
    \textbf{c}^{(AT)}&={\sum_{i}} \mathbf{sg}(\alpha^{(A)}_i)~{\textbf{H}^{(T)}_i},
\end{align}
where $\textbf{q}^{(T)}$ and $\textbf{q}^{(A)}$ are the global queries used to decide which parts of the aligned signals to focus on based on each modality perspective.
Additionally, $\textbf{c}^{(xy)}$s are context vectors, where the $x$ represents the modality used as a query and the $y$ represents the modality used as a key-value pair.
To prevent the CAN from learning attention based on the other modality, we introduce a function $\mathbf{sg}$; stop-gradient operator as shown in the equations (7) and (9).
It cuts the gradient flow through its argument during the backpropagation.

\subsection{Training objective}
In the training, the CAN makes three different predictions using the context vectors.
\begin{gather}
    \hat{y}= \text{softmax}([{c}^{(TT)};{c}^{(TA)};{c}^{(AA)};{c}^{(AT)}])^\intercal~\textbf{W}+\textbf{b}), \\[10pt]
    \hat{y}^{(T)}= \text{softmax}(({c}^{(TT)})^\intercal~\textbf{W}^{(T)}+\textbf{b}^{(T)}~),\\[10pt]
    \hat{y}^{(A)}= \text{softmax}(({c}^{(AA)})^\intercal~\textbf{W}^{(A)}+\textbf{b}^{(A)}~),
\end{gather}
where the $\textbf{W}$s and $\textbf{b}$s are trainable weights.
$\hat{y}$ is made based on all the context vectors and each $\hat{y}^{(T)}$ and $\hat{y}^{(A)}$ is made based on a context vector that uses either the text or the audio modality.
Using the predictions, we calculate loss terms as follows:
\begin{gather}
    \mathcal{L}_{align}= CE(\hat{y}^{(T)},~y) + CE(\hat{y}^{(A)},~y),\\[10pt]
    \mathcal{L}_{total}= CE(\hat{y},~y) + \alpha \cdot \mathcal{L}_{align},
\end{gather}
where CE represents the cross-entropy loss, $y$ is the true emotion labels, and $\alpha$ is a weight for the additional loss term $\mathcal{L}_{align}$, of which optimal value is found using the validation dataset.
The additional loss terms in $\mathcal{L}_{align}$ are added to help the global attention attend to the salient features based on each modality better.
\begin{gather}
    \hat{y}^\text{final}=(\hat{y})\cdot(\hat{y}^{(T)})^\alpha \cdot(\hat{y}^{(A)})^\alpha
\end{gather}
After the training, the final prediction $\hat{y}^\text{final}$ is calculated following the equation (15).

\section{Experiments}
In this section, we describe the experimental setup and the results conducted on the IEMOCAP dataset.
First, we compare the CAN to other SER models for the weighted accuracy (\textbf{WA}) and the unweighted accuracy (\textbf{UA}), where the CAN shows the best performance.
In addition, we conduct several analyses on our model to see how each component described in Section \ref{section:algorithm} affects the performance of the CAN.

\subsection{Dataset}
In the experiments, we use the Interactive Emotional Dyadic Motion Capture (IEMOCAP) \cite{busso2008iemocap} dataset which provides the speech and text dataset including the alignment information as represented in Table \ref{tab:alignment}.
%The information is obtained using conventional acoustic speech models from the Sphinx-\Romannum{3} (version 3.0.6).
Each utterance in the dataset is labeled as one of the 10-class emotions, where we do not use the classes with too few data instances (fear, disgust, other) so the final dataset contains 7,486 utterances in total (1,103 angry, 1,040 excite, 595 happy, 1,084 sad, 1,849 frustrated, 107 surprise and 1,708 neutral).
In the experiments, we perform 10-fold cross-validation, and in each validation, the total dataset is split into 8:1:1 training set, validation set, and test set, respectively.

\subsection{Experimental setup}
\label{subsection:setup}
For text input, we use a sequence of words in Table \ref{tab:alignment} as our text input and the 300-dimensional GloVe word vectors \cite{pennington2014glove} are used as the embedding vectors.
In this step, we remove the special tokens such as `$\left< s \right>$', `$\left< sil \right>$', `$\left< /s \right>$' and their durations are equally divided into the neighboring words.
For audio input, we use the zero-padded MFCC segments as our audio input, which are obtained as described in Section \ref{subsection:audioprocessing}.
In the MFCC conversion, we use 40 MFCC coefficients and the frames are extracted while sliding the hamming window with 25ms frame size and 10ms hopping.
We use the bidirectional LSTMs with 128 hidden units followed by the dropout layer with 0.3 dropout probability.
For the cross attention, multi-head global attention with four heads is used to view the inputs from various perspectives so enrich the aggregated information \cite{vaswani2017attention}.
During the training, we use the validation dataset as a criterion of early stopping with the patience 10.
We use the batch size of 64 and use the Adam optimizer \cite{kingma2014adam} with a learning rate of 1e-3, and the gradients are clipped with a norm value of 1.0.
The weight of the additional loss term $\alpha$ is set to 0.1, which is obtained from the cross-validation.

\subsection{Results}
\subsubsection{Performance comparison}
\begin{table}[h]
  \caption{Comparison of the models. The accuracy values are represented as an average of the 10-fold validations and the standard deviations are written next to them.}
  \label{tab:accuracy}
  \centering
  \begin{tabular}
    {L{0.3\columnwidth}C{0.25\columnwidth}C{0.25\columnwidth}}
    \toprule
    \multicolumn{1}{c}{\textbf{Model}} & \textbf{WA} & \textbf{UA} \\
    \midrule
    TextModel \cite{mirsamadi2017automatic}  &   0.513  {\scriptsize $\pm$ 0.015}  &   0.443 {\scriptsize $\pm$ 0.015} \\
    AudioModel \cite{mirsamadi2017automatic} &   0.431 {\scriptsize $\pm$ 0.017}  &   0.323 {\scriptsize $\pm$ 0.015} \\    
    Yoon et al. \cite{yoon2019speech}  &   0.564 {\scriptsize $\pm$ 0.020} &   0.472 {\scriptsize $\pm$ 0.017}  \\
    Xu et al. \cite{xu2019learning} &   0.560 {\scriptsize $\pm$ 0.028} &   0.450 {\scriptsize $\pm$ 0.028}   \\
    \midrule
    CAN (ours) &   \textbf{0.579} {\scriptsize $\pm$ \textbf{0.019}} &   \textbf{0.487} {\scriptsize $\pm$ \textbf{0.017}}   \\
    \bottomrule
    \end{tabular}
\end{table}

% \midrule
%         \\
Table~\ref{tab:accuracy} shows the performance of the CAN and the other SER models. Each `TextModel' and `AudioModel' uses a single modality by encoding it with a simple bidirectional LSTM with the global attention following \cite{mirsamadi2017automatic}. The other two multimodal models are \cite{yoon2019speech} and \cite{xu2019learning} proposed in the previous studies, where the attention weights are obtained based on the interaction between the audio and the text modalities.
In the experiments, we re-implement all the models and obtain the accuracy values as described in Section \ref{subsection:setup}.
As the Table~\ref{tab:accuracy} shows, our CAN outperforms the other models for both \textbf{WA} and \textbf{UA} including the previous state-of-the-art model \cite{yoon2019speech}.
To analyze the causes of the performance gain, we conduct further experiments to see the effectiveness of the components in our methodology, which are described in the next sections.
% \textcolor{red}{The results support our assumption that obtaining information from the aligned audio and text is effective in SER.}
% \textcolor{red}{The results support our assumption that obtaining information from the aligned audio and text without learning the complex attention between different modalities is effective in SER.}
%The confusion matrix of the CAN is drawn in Figure \ref{fig:confusion}

\begin{table}[h]
  \caption{Comparison of the segmentation policies.}
  \label{tab:segmentation}
  \centering
  \begin{tabular}
    {L{0.37\columnwidth}C{0.23\columnwidth}C{0.23\columnwidth}}
    \toprule
    \multicolumn{1}{c}{\textbf{Segmentation}} & \textbf{WA} & \textbf{UA} \\
    \midrule
    Aligned   &   \textbf{0.579} {\scriptsize $\pm$ 0.019}  &   \textbf{0.487} {\scriptsize $\pm$ \textbf{0.017}} \\
    Equal &   0.568 {\scriptsize $\pm$ 0.022}  &   0.467 {\scriptsize $\pm$ 0.021} \\    
    \bottomrule
    \end{tabular}
\end{table}
\subsubsection{Segmentation policy}
In order to demonstrate the superiority of the aligned segmentation, we compare it to the segmentation where a 1-dimensional audio signal is segmented into the segments of equal length, which has been widely used in the previous studies \cite{sahoo2019segment, mao2019deep}.
% The equal segmentation method has been widely used in the previous studies \cite{sahoo2019segment, mao2019deep}.
In the experiment, the aligned segmentation outperforms the equal segmentation for both \textbf{WA} and \textbf{UA}.
The results in Table \ref{tab:segmentation} imply that our aligned segmentation actually has effectiveness in combining the information in the cross attention.

\subsubsection{Ablation study}
\begin{table}[h]
  \caption{Accuracy comparison in the ablation studies}
  \label{tab:ablation}
  \centering
  \begin{tabular}
    {L{0.37\columnwidth}C{0.23\columnwidth}C{0.23\columnwidth}}
    \toprule
    \multicolumn{1}{c}{\textbf{Model}} & \textbf{WA} & \textbf{UA} \\
    \midrule
    CAN   &   \textbf{0.579} {\scriptsize $\pm$ \textbf{0.019}}  &   \textbf{0.487} {\scriptsize $\pm$ \textbf{0.017}} \\
    \midrule
    - stop-gradient  &   0.570 {\scriptsize $\pm$ 0.018}  &   0.469 {\scriptsize $\pm$ 0.023} \\
    - $\mathcal{L}_{align}$  &   0.573 {\scriptsize $\pm$ 0.015}  &   0.484 {\scriptsize $\pm$ 0.017} \\
    - stop-gradient, $\mathcal{L}_{align}$ &  0.563 {\scriptsize $\pm$ 0.015}  &   0.479 {\scriptsize $\pm$ 0.024} \\
    - cross attention &   0.556 {\scriptsize $\pm$ 0.013} &   0.458 {\scriptsize $\pm$ 0.015} \\
    \bottomrule
    \end{tabular}
\end{table}

As supportive evidence of the components in our model, we conduct ablation studies for four variant models while removing each of the components in Section \ref{section:algorithm}.
When we remove the stop-gradient operator and additional loss term $\mathcal{L}_{align}$, the accuracy of the CAN decreases and the worst performance for the \textbf{WA} is observed when both components are removed.
Furthermore, we even remove the whole cross attention in the CAN, where the prediction is based only on a concatenation of the $c^{\text{(TT)}}$ and $c^{\text{(AA)}}$ and the stop-gradient operator and the $\mathcal{L}_{align}$ are not used.
In that case, the performance decreases even more compared to the other variants.

\section{Conclusion}
% In this paper, we propose a Cross Attention Network (CAN) that uses the cross attention to combine information from the aligned audio and text signals.
% Inspired by the way humans recognize speech, we align the text and audio signals using aligned segmentation, while enabling the CAN to focus on different the salient parts of the speech based on each modality perspective.
% In the experiments conducted on the IEMOCAP dataset, the CAN outperforms the state-of-the-art system and the ablation studies show that the superiority of the CAN is actually affected by its components such as aligned segmentation, stop-gradient operator, and the additional loss.
% 어떤 효과를 기대하는지, 어떤 응용에 쓰일 수 있을 지, future work로는 어떤 것들이 있을지

In this paper, we propose a Cross Attention Network (CAN) for Speech Emotion Recognition (SER) task.
It uses the cross attention to combine information from the aligned audio and text signals.
Inspired by the way humans recognize speech, we align the text and audio signals so that the CAN regards each modality to have the same time resolution.
In the experiments conducted on the IEMOCAP dataset, the proposed system outperforms the state-of-the-art systems by 2.66\% and 3.18\% relatively for the weighted and unweighted accuracy.
% \textcolor{red}{In the future work, we plan to integrate the automatic speech recognition (ASR) system with our method. 
% Since the ASR system outputs the text and alignment information together, it will be possible to apply our system to the real-world scenario where only the speech signal without the text and the alignment information is provided.}
To the best of our knowledge, this is the first study that shows the improvement
using the aligned audio and text signals in SER.
In order to apply our system to the real-world scenario where only the speech signal is available, the text and alignment information are required for the CAN to work properly.
In future work, we plan to extend our research by integrating the CAN with the automatic speech recognition system which outputs the text and alignment information given a speech signal.

\section{Acknowledgments}
K. Jung is with ASRI, Seoul National University, Korea.
This work was supported by the Ministry of Trade, Industry \& Energy (MOTIE, Korea) under the Industrial Technology Innovation Program (No. 10073144).
This research was results of a study on the ``HPC Support'' Project, supported by the `Ministry of Science and ICT' and NIPA.

% No ack at the submission stage
% \input{7acknowledge.tex}

\bibliographystyle{IEEEtran}
\bibliography{0Interspeech2020}

% Generated by IEEEtran.bst, version: 1.13 (2008/09/30)
\begin{thebibliography}{10}
\providecommand{\url}[1]{#1}
\csname url@samestyle\endcsname
\providecommand{\newblock}{\relax}
\providecommand{\bibinfo}[2]{#2}
\providecommand{\BIBentrySTDinterwordspacing}{\spaceskip=0pt\relax}
\providecommand{\BIBentryALTinterwordstretchfactor}{4}
\providecommand{\BIBentryALTinterwordspacing}{\spaceskip=\fontdimen2\font plus
\BIBentryALTinterwordstretchfactor\fontdimen3\font minus
  \fontdimen4\font\relax}
\providecommand{\BIBforeignlanguage}[2]{{%
\expandafter\ifx\csname l@#1\endcsname\relax
\typeout{** WARNING: IEEEtran.bst: No hyphenation pattern has been}%
\typeout{** loaded for the language `#1'. Using the pattern for}%
\typeout{** the default language instead.}%
\else
\language=\csname l@#1\endcsname
\fi
#2}}
\providecommand{\BIBdecl}{\relax}
\BIBdecl

\bibitem{kolakowska2014emotion}
A.~Ko{\l}akowska, A.~Landowska, M.~Szwoch, W.~Szwoch, and M.~R. Wrobel,
  ``Emotion recognition and its applications,'' in \emph{Human-Computer Systems
  Interaction: Backgrounds and Applications 3}.\hskip 1em plus 0.5em minus
  0.4em\relax Springer, 2014, pp. 51--62.

\bibitem{nwe2003speech}
T.~L. Nwe, S.~W. Foo, and L.~C. De~Silva, ``Speech emotion recognition using
  hidden markov models,'' \emph{Speech communication}, vol.~41, no.~4, pp.
  603--623, 2003.

\bibitem{chavhan2010speech}
Y.~Chavhan, M.~Dhore, and P.~Yesaware, ``Speech emotion recognition using
  support vector machine,'' \emph{International Journal of Computer
  Applications}, vol.~1, no.~20, pp. 6--9, 2010.

\bibitem{mao2014learning}
Q.~Mao, M.~Dong, Z.~Huang, and Y.~Zhan, ``Learning salient features for speech
  emotion recognition using convolutional neural networks,'' \emph{IEEE
  transactions on multimedia}, vol.~16, no.~8, pp. 2203--2213, 2014.

\bibitem{mirsamadi2017automatic}
S.~Mirsamadi, E.~Barsoum, and C.~Zhang, ``Automatic speech emotion recognition
  using recurrent neural networks with local attention,'' in \emph{2017 IEEE
  International Conference on Acoustics, Speech and Signal Processing
  (ICASSP)}.\hskip 1em plus 0.5em minus 0.4em\relax IEEE, 2017, pp. 2227--2231.

\bibitem{castellano2008emotion}
G.~Castellano, L.~Kessous, and G.~Caridakis, ``Emotion recognition through
  multiple modalities: face, body gesture, speech,'' in \emph{Affect and
  emotion in human-computer interaction}.\hskip 1em plus 0.5em minus
  0.4em\relax Springer, 2008, pp. 92--103.

\bibitem{yoon2020attentive}
S.~Yoon, S.~Dey, H.~Lee, and K.~Jung, ``Attentive modality hopping mechanism
  for speech emotion recognition,'' in \emph{ICASSP 2020-2020 IEEE
  International Conference on Acoustics, Speech and Signal Processing
  (ICASSP)}.\hskip 1em plus 0.5em minus 0.4em\relax IEEE, 2020, pp. 3362--3366.

\bibitem{yoon2019speech}
S.~Yoon, S.~Byun, S.~Dey, and K.~Jung, ``Speech emotion recognition using
  multi-hop attention mechanism,'' in \emph{ICASSP 2019-2019 IEEE International
  Conference on Acoustics, Speech and Signal Processing (ICASSP)}.\hskip 1em
  plus 0.5em minus 0.4em\relax IEEE, 2019, pp. 2822--2826.

\bibitem{xu2019learning}
H.~Xu, H.~Zhang, K.~Han, Y.~Wang, Y.~Peng, and X.~Li, ``Learning alignment for
  multimodal emotion recognition from speech,'' \emph{arXiv preprint
  arXiv:1909.05645}, 2019.

\bibitem{raffel2017online}
C.~Raffel, M.-T. Luong, P.~J. Liu, R.~J. Weiss, and D.~Eck, ``Online and
  linear-time attention by enforcing monotonic alignments,'' in
  \emph{Proceedings of the 34th International Conference on Machine
  Learning-Volume 70}.\hskip 1em plus 0.5em minus 0.4em\relax JMLR. org, 2017,
  pp. 2837--2846.

\bibitem{battenberg2019location}
E.~Battenberg, R.~Skerry-Ryan, S.~Mariooryad, D.~Stanton, D.~Kao, M.~Shannon,
  and T.~Bagby, ``Location-relative attention mechanisms for robust long-form
  speech synthesis,'' \emph{arXiv preprint arXiv:1910.10288}, 2019.

\bibitem{bertero2017first}
D.~Bertero and P.~Fung, ``A first look into a convolutional neural network for
  speech emotion detection,'' in \emph{2017 IEEE international conference on
  acoustics, speech and signal processing (ICASSP)}.\hskip 1em plus 0.5em minus
  0.4em\relax IEEE, 2017, pp. 5115--5119.

\bibitem{sahoo2019segment}
S.~Sahoo, P.~Kumar, B.~Raman, and P.~P. Roy, ``A segment level approach to
  speech emotion recognition using transfer learning,'' in \emph{Asian
  Conference on Pattern Recognition}.\hskip 1em plus 0.5em minus 0.4em\relax
  Springer, 2019, pp. 435--448.

\bibitem{sebastian2019fusion}
J.~Sebastian and P.~Pierucci, ``Fusion techniques for utterance-level emotion
  recognition combining speech and transcripts,'' in \emph{Proc. Interspeech},
  2019, pp. 51--55.

\bibitem{liang2019cross}
J.~Liang, S.~Chen, J.~Zhao, Q.~Jin, H.~Liu, and L.~Lu, ``Cross-culture
  multimodal emotion recognition with adversarial learning,'' in \emph{ICASSP
  2019-2019 IEEE International Conference on Acoustics, Speech and Signal
  Processing (ICASSP)}.\hskip 1em plus 0.5em minus 0.4em\relax IEEE, 2019, pp.
  4000--4004.

\bibitem{pennington2014glove}
J.~Pennington, R.~Socher, and C.~Manning, ``Glove: Global vectors for word
  representation,'' in \emph{Proceedings of the 2014 conference on empirical
  methods in natural language processing (EMNLP)}, 2014, pp. 1532--1543.

\bibitem{hochreiter1997long}
S.~Hochreiter and J.~Schmidhuber, ``Long short-term memory,'' \emph{Neural
  computation}, vol.~9, no.~8, pp. 1735--1780, 1997.

\bibitem{busso2008iemocap}
C.~Busso, M.~Bulut, C.-C. Lee, A.~Kazemzadeh, E.~Mower, S.~Kim, J.~N. Chang,
  S.~Lee, and S.~S. Narayanan, ``Iemocap: Interactive emotional dyadic motion
  capture database,'' \emph{Language resources and evaluation}, vol.~42, no.~4,
  p. 335, 2008.

\bibitem{vaswani2017attention}
A.~Vaswani, N.~Shazeer, N.~Parmar, J.~Uszkoreit, L.~Jones, A.~N. Gomez,
  {\L}.~Kaiser, and I.~Polosukhin, ``Attention is all you need,'' in
  \emph{Advances in neural information processing systems}, 2017, pp.
  5998--6008.

\bibitem{kingma2014adam}
D.~P. Kingma and J.~Ba, ``Adam: A method for stochastic optimization,''
  \emph{arXiv preprint arXiv:1412.6980}, 2014.

\bibitem{mao2019deep}
S.~Mao, P.~Ching, and T.~Lee, ``Deep learning of segment-level feature
  representation with multiple instance learning for utterance-level speech
  emotion recognition,'' \emph{Proc. Interspeech 2019}, pp. 1686--1690, 2019.

\end{thebibliography}

\end{document}